\documentclass[12pt]{iopart}
\usepackage{iopams}
\def\({\left(}
\def\){\right)}
\def\[{\left[}
\def\]{\right]}
\def\be{\begin{eqnarray}}
\def\ee{\end{eqnarray}}
\def\ne{\nonumber\end{eqnarray}}
\def\d{{\cal D}}

\def\E{{\mathcal{E}}}

\def\exp#1{\mathop{\rm exp}\nolimits \left\{#1\right\}}

\def\Det{\mathop{\rm Det}\nolimits}
\def\Ln{\mathop{\rm Ln}\nolimits}
\def\Ln{\mathop{\rm Ln}}

\def\d{{\cal D}}

\begin{document}

\title[P-odd cylindrical shell in the framework of QED]
    {Parity violating cylindrical shell in the framework of QED}

\author{I V  Fialkovsky\dag, V N  Markov \P \ and  Yu M Pis'mak\dag}
\address
{\dag \ Department of Theoretical Physics,  State University of
Saint-Petersburg, Russia}
\address
{\P \ Department of Theoretical Physics, Petersburg Nuclear
Physics Institute, Russia}
\ead{\dag ignat.fialk@paloma.spbu.ru, pismak@jp7821.spb.edu, \P markov@thd.pnpi.spb.ru, }

\begin{abstract}
We present calculations of Casimir energy (CE) in a system
of quantized electromagnetic (EM) field
interacting with an infinite circular cylindrical shell (which we call `the defect').
%
%
Interaction is described in the only QFT-consistent
way by Chern-Simon action concentrated on the defect,
with a single coupling constant~$a$.

For regularization of UV divergencies of the theory we use 
Pauli-Villars regularization of the free EM action. The divergencies are extracted
as a polynomial in regularization mass $M$,
and they renormalize classical part of the surface action.

We reveal the dependence of CE on the coupling constant $a$. Corresponding Casimir force is
attractive for all values of $a$. For $a\rightarrow\infty$ we reproduce the
known results for CE for perfectly conducting cylindrical shell
first obtained by DeRaad and Milton.
%
%

As a future task for solving existing arguments on observational
status of (rigid) self-pressure of a single object, we propose for
investigation a system which we call `Casimir drum'.

\end{abstract}


\submitto{\JPA}


\section{Introduction}
Predicted in 1948 \cite{Casimir'48} Casimir effect has been for a long
time a non-detectable theoretical `play of mind'. Prophesying an
attractive force between two neutral parallel conducting planes in
vacuum, the Casimir effect is a pure quantum one --- there is no
such force between the planes in classical electrodynamics.
%

Development of the experimental technique allowed first to observe
Casimir effect \cite{Sparnaay 58} and then to measure it
with 5-10\% accuracy \cite{Lamoreaux 97}. Nowadays, due to works
of Mohideen and his colleagues, as well as many other groups, the
total experimental error for the force between metal surfaces is
reduced to 0.5\% \cite{Mohideen 98}-\cite{Klimchitskaya 05}.

Original Casimir's configuration of parallel (perfectly
conducting) planes is well studied both theoretically and
experimentally. Still there are considerable lacunas in our
understanding of the effect for complicated geometries  and
non-perfect materials. The particular interest for cylindrical
geometry which we consider in this paper, is motivated by rapid
development of the carbon nanotubes technology. For discussion of
possible role of the Casimir force in dynamics and stability
of micro- and nano- electromechanical devices (MEMS and NEMS) see
\cite{Barcenas'05, Lin Zhao'07}, and reference therein.


From the theoretical point of view, the Casimir force between distinct bodies
is well established, especially taking into
account the main achievements (both analytical and numerical) of the recent time
\cite{Emig Jaffe 06}-\cite{Kenneth 07}.
On the other hand, calculations of the Casimir effect for a single object
(self-energy, self-stress, etc.)
still provoke controversies \cite{Jaffe 02-04, Milton 0401},
and a self-consistent description of systems
with sharp material boundaries
is yet to be developed in the framework of the quantum field theory (QFT).
In this paper we address both of these issues.

The paper is organized as following. In Part 2 we discuss the construction of the model
for the system of quantum electrodynamics' (QED)
fields interacting with the cylindrical shell. In Part 3 we sketch
our approach to calculation of the modified propagator of the system and the Casimir energy,
and give their explicit form.
In Part 4 we present the conclusions and discuss the perspectives of the work.

\section{Statement of the problem}
There are several ways to model the presence of
matter (which we also call `spatial defect') in QFT. The simplest one
is to fix the values of quantum fields and/or their derivatives with
boundary conditions (BC) on the surface of the defect.
%
%
However, imposing BC is physically unjustified as it constraints all modes of the fields.
At the same time, in reality field's modes with  high enough frequencies propagate
freely through any material boundary.

The most natural generalization of BC is to couple the quantum fields to classical external
field (background) supported spatially on the defect.
The simplest case of such background is a singular one with delta-function profile.
Its introduction into the classical action is equivalent to
imposing  matching conditions  on the quantum fields
which model semitransparent boundaries.
Delta-potential is an effective way to describe a thin film present in the system,
when its thickness is negligible compared to the distances in the range of interest.
%
%
In a certain limit (usually the strong coupling one) delta-potentials reproduce simplest
BC such as Dirichlet, Neumann or Robin ones in the case of scalar fields \cite{Symanzik'81}.
%

In the context of QFT such interaction of quantum fields
with delta-potentials (introduced as a part of the action) must be constructed
satisfying the basic principles of the theory --- locality, gauge and Lorenz invariance,
renormalizability. For the first time this issue was addressed and
thoroughly studied (for the case of massless scalar fields) by
Symanzik \cite{Symanzik'81} in 1981. Since then, there were a number of
Casimir calculations with delta-potentials, see for instance
\cite{Bordag'98}-\cite{Milton 02}.
However, until very lately the issue of renormalizability of a theory with delta-potential
still invoked contradictions \cite{Jaffe 02-04, Milton 0401}, and was apparently resolved in
\cite{Bordag 04}.
Still all of the existing papers deal only with simplified scalar models, usually in lower dimensions.
In a limited number of particular cases
scalar fields can be combined as TM and TE modes \cite{Stratton} of EM field,
to describe some specific aspects of Casimir problems in QED. Until
\cite{Markov Pismak'05,Fialkovsky 03-06} there were no attempts to construct
a self-consistent QED model with a delta-potential interaction satisfying all QFT principles
and allowing one to describe self-consistently all possible observable consequences of the presence
of a defect.
In this paper we generalize the results
of \cite{Markov Pismak'05} to the case of cylindrical geometry.

Following the approach of \cite{Markov Pismak'05}, we construct a
QED model with photon field coupled to the defect through
a delta-potential supported on the surface of an infinite circular cylinder.
%
%
%
%
%
We neglect interaction of fermion fields with the defect since
any observable consequences of such interaction
are exponentially suppressed at the distances larger then
inverse electron mass $m^{-1}_e\approx10^{-10}$cm \cite{Fialkovsky 03-06,Fialkovsky 06}.
Thus, massive fermion fields cannot contribute to the Casimir force which
has macroscopical (experimentally verified) values at the scale of $10$--$100 nm$.
We can neglect Dirac fields and consider pure photodynamics.

For the photon field $A_\mu$ and defect surface described by
equation $\Phi(x)=0$ we construct the action as a sum
 \be
   S= S_{0}+ S_{def}
    \label{S}
 \ee
of usual Maxwell action of electromagnetic field (throughout the paper we set $c=\hbar=1$)
 \be
    S_{0}=-\frac{1}{4} \int d^4 x\,F_{\mu\nu}(x)F^{\mu\nu}(x)
    ,\qquad  F_{\mu\nu}(x)=\partial_{\mu}A_{\nu}-\partial_{\nu}A_{\mu}
    \label{S_EM}
\ee
and defect action of Chern-Simon type \cite{Chern'71}
\begin{eqnarray}
S_{def}=\frac{a}{2} \int{ d^{4}x
\varepsilon^{\mu\nu\rho\sigma}\partial_{\mu}\Phi(x)
    \delta(\Phi(x))A_{\nu}(x)F_{\rho\sigma}(x)}\label{S_def}.
\end{eqnarray}
Here $\varepsilon_{\mu\nu\rho\sigma}$ is totaly antisymmetric
tensor ($\varepsilon_{0123}=1$), $a$ --- dimensionless coupling constant.
For a cylinder of radius $R$ placed along the $x_3$ axis   we have
\be
    \Phi(x)=x_1^2+x_2^2-R^2.
     \label{Phi}
\ee
We must stress here that the form of the defect action (\ref{S_def})
is completely determined by above mentioned basic principles of QFT.
In particular, introduction of
any other local, gauge and lorenz invariant terms (with higher derivatives, etc.)
unavoidably brings
to the theory coupling constants of negative dimensions. Such
theories have infinite number of primitively divergent diagrams,
being unrenormalizable in conventional sense \cite{Symanzik'81,
Collins'84}. Thus we are left with the Chern-Simon defect action
that is space parity violating and this unusual property quite
naturally arises at the very beginning of our consideration. We
will show below that in the limit of $a\to\infty$ the Casimir energy for perfectly
conducting cylindrical shell is reproduced.

\section{Casimir energy and photon propagator}
All properties of a QFT system can be described
if its generating functional is known
  \be
G(J)={\cal N}\int\d A\,\exp{i{\cal S}+JA}.
    \label{G(J)}
 \ee
To render the theory finite, instead of $S$ (\ref{S}) we set into $G(J)$
\be
    {\cal S}=S_0^{reg}+S_{def}
    \label{S_reg}
\ee
where we introduced Pauli-Villars \cite{Pauli Vilars'49}
UV regularization of the theory
%
%
%
\be
S^{reg}_0=-\frac{1}{4}F_{\mu\nu}F^{\mu\nu}-
   M^{-2}\partial_\lambda F_{\mu\nu} \partial^\lambda F^{\mu\nu}
\ee
%
%
here and below we omit the sign of integration, $S_{def}$ is defined in (\ref{S_def}).
In the limit $M\to\infty$ we return to action $S$ (\ref{S}),
but then the theory posses both standard UV divergencies of QED and
specific (geometry dependent) ones in the vacuum loops --- in Casimir energy in particular.
These divergences can be canceled by counter-terms in the framework of renormalization procedure.
If $M$ is finite there are no divergences in the model with the action ${\cal S}$.

For definition of the normalization constant ${\cal N}$ in (\ref{G(J)}) we use the following condition
\be
    G(0)|_{a=0}=1 \quad\Longrightarrow\quad
        {\cal N}^{-1}= \int\d A\,\exp{i{{\cal S}\big|_{a=0}}}
    \label{N}
\ee
which means that in pure photodynamics without a defect $\ln G(0)$ vanishes. This sets the
reference point for the values of
the energy density (per unit length of the cylinder)
of the system as the later one is expressed through  the value  of
$G(0)$
\be
    \E =-\frac1{iT L} \ln G(0), \qquad T=\int dx_0,\quad L=\int dx_3.
    \label{E}
\ee
%
%
%

For explicit calculations of $G(J)$ we first
 choose the coordinate basis associated with cylindrical geometry
and transform the vectors accordingly
$$
    A=(A_0,A_1,A_2,A_3)\quad\rightarrow \quad \bar{A}\equiv \tau A =(A_0,A_\bot,A_{\|},A_3)
$$
with
\be
{\rm \tau}=\(\matrix{
    1&0&0&0\cr
    0&\cos\phi&\sin\phi&0\cr
    0&-\sin\phi&\cos\phi&0\cr
    0&0&0&1}\),
    \label{tau}
 \ee
where $\phi$ is polar angle in $x_1 x_2$ plane.
After such linear transformation of integration variables in (\ref{G(J)})
the defect action (\ref{S_def})
does not contain anymore neither vector components $A_\bot$ nor derivatives $\partial_\bot$
transversal to the defect.
%
%
This makes it possible to represent the defect action contribution as an 
integral over auxiliary fields defined on the defect surface only.
%
%
Then functional integration becomes purely Gaussian and for $G(J)$
one calculates explicitly
\be
    G(J)=(\Det {Q})^{-1/2}\exp{\frac12 { J}\( \mathcal{D}+\tau^T\bar \mathcal{H}\tau\){ J} }.
    \label{Z_QS}
\ee
Here $\mathcal{D}\equiv \mathcal{D}_{\mu\nu}(x)$
is the standard (UV regularized) free photon propagator, $\bar \mathcal{H}$
defines its corrections due to the presence of the defect,
and $Q$ describes the dependence of the Casimir energy on the geometry of the defect.
In Fourier-components respecting the symmetries of the system the propagator  in Feynman gauge
can be written as
\be
    \mathcal{D}_{\mu\nu}(x,y)=\delta_{\mu\nu}\int \frac{dp_S} {(2\pi)^3} e^{ip_S (x_S-y_S)}
         D(p_S; \rho_x,\rho_y)
\ee
$$
    D(p_S; \rho_x,\rho_y)=I_n(i|p_S|\rho_<) K_n(i|p_S|\rho_>)
        -I_n(q_S\rho_<) K_n(q_S\rho_>), \qquad q_S=\sqrt{M^2-p_S^2}
$$
where $x_S=(x_0,\phi,x_3)$, $p_S=(p_0,n,p_3)$, $\int dp_S=\int dp_0 dp_3 \sum_{n=-\infty}^{\infty}$,
$I_n$, $K_n$ are modified Bessel functions, $\rho_{<,>}$ is smaller (bigger)
of $\rho_{x,y}$, $\rho_{x}=\sqrt{x_1^2+x_2^2}$ --- polar radius in the plane $x_1x_2$.
In the same Fourier representation for $\bar \mathcal{H}(x,y)$ and $Q(x_S,y_S)$ we can write
\begin{eqnarray}
    \bar \mathcal{H}_{ab}(p_S;\rho_x,\rho_y) =  a D(p_S;\rho_x,R) L_{ac}Q^{-1}_{cb}(p_S)D(p_S;R,\rho_y)\\
    Q_{ab}(p_S)=\delta_{ab}+a D(p_S; R,R) L_{ab},
        \label{Q_ab}\\
    L_{ab}=2 i R \varepsilon_{abc}p_S^c,\qquad a,b,c=1,2,3
\end{eqnarray}
%
%

Using the famous $\tr \ln = \ln \det$ identity
one can express the Casimir energy (\ref{E}) as
\be
  \E=\frac1{2iTL}\int dx_S  \tr (\Ln Q(x_S-y_S))_{x_S=y_S}
    =\frac{1}{8 \pi^2}\int d p_S \,\tr \ln Q(p_S)
  \label{tr ln}
\ee
where $\tr \ln Q(p_S)$ denotes the sum of diagonal elements of the
$3\times 3$-matrix $\ln Q(p_S)$.
Putting (\ref{Q_ab}) into (\ref{tr ln}) one writes for the energy density
\be
\E ={1\over 4\pi R^2}\int_0^{\infty} pdp
   \sum_{n=-\infty}^{\infty} \ln\Big(1+ a^2 Y_n^M(p)\Big)
   \label{E/L}
\ee
\be
Y_n^M(p)= -4 \(I_n(p) K_n(p)-I_n'(q) K_n(q)\)\times
\ne
\be
\qquad
\times\(
    p^2 \(I_n'(p) K_n'(p)-  I_n'(q) K_n'(q)\)
    + {n^2 (RM)^2\over p^2+(RM)^2}I_n(q) K_n(q)
    \)
\ne
where
$q=\sqrt{p^2+(RM)^2}$.

It is easy to check that (\ref{E/L}) is finite for any fixed
value of auxiliary mass $M$ and diverges in the limit $M\to\infty$.
With help of uniform (Debye) asymptotics of the Bessel functions \cite{Abram}
we subtract the most divergent terms
from the integrant to make it finite when regularization is removed, and
add them explicitly. Following then the renormalization procedure we
extract from the substraction exact (polynomial and/or logarithmic)
dependence on $M$, and construct the
counter terms. This is done with help of generalized Abel-Plana formula \cite{Saharian}
and its modification~\cite{Fialkovsky 07}.

As a result of the calculations we present the energy density (\ref{E/L}) as
a sum
\be
    \E=\E_{Cas}+\Delta
\ee
of finite Casimir energy
\be
    \E_{Cas}=\frac{1}{4\pi R^2}\(\frac{a^2 \ln 2\pi}{4(1+a^2)}
        +\int_0^\infty pdp\,\, \epsilon^{finite}\)
        \label{E_cas}\\
\epsilon^{finite}=\ln\(\frac{1+ a^2 Y_0(p)}{1+a^2}\)
    +\frac{a^2 }{1+a^2}\frac{p^4 }{4(1+p^2)^3}
\nonumber\\ \qquad
        + 2\sum_{n=1}^{\infty}  \(
        \ln\(\frac{1+ a^2 Y_n(p)}{1+a^2}\)
            +\frac{ a^2 }{1+a^2}\frac{p^4 }{4(n^2+p^2)^3}\)
\ne
here $ Y_n(p)=\lim_{M\to\infty} Y_n^M(p)= - 4 p^2 I_n(p)K_n(p)I_n'(p)K_n'(p)$,
and the counter-terms
\be
\Delta =
     R M^3 A_3  +\frac{M}{R} A_1
    \label{Delta}
\ee
where
$$
    A_1=\frac{2 a^2}{\pi(1+a^2)}\int_0^\infty pdp \int_0^\infty dn\  y_n^{(1)}(p),\quad
    A_3=\frac1{2 \pi} \int_0^\infty pdp \int_0^\infty dn  \ln\(1+a^2 y_n^{(0)}(p)\),
$$
with
$$
y_n^{(0)}=2- (n^2+p^2+1)^{-1}
    -{2\sqrt{n^2+p^2}}/{\sqrt{n^2+p^2+1}}
$$
and $y_n^{(1)}(p)$ is the first order term of uniform asymptotic of $Y_n(p)|_{M=1}$.

$\E_{Cas}$ as function of $a$ is real for $a\in\Re$ and for $a\in i(-1,1)$.
The later region is out of physical interest \cite{Scandurra'00}
as for this case the action (\ref{S}), (\ref{S_reg})
acquires imaginary part.
%
%
For all physically sensible values of $a$, $\E_{Cas}$ is positive giving rise to
attractive Casimir force.
In the limit $a\to\infty$, one can easily derive that (\ref{E_cas}) coincides explicitly
with known results for
the Casimir energy of a perfectly conducting cylinder \cite{cond cylinder}.

For renormalization of the counter-terms (\ref{Delta})
we must introduce into the action (\ref{S_reg})
also the classical energy density
\be
    E=R \sigma_0 + \frac{h_0}R
\ne
where bare parameters $\sigma_0$, $h_0$ are the surface tension
and inverse radius parameter correspondingly.
To renormalize the divergencies we make following redefinition
\be
    \sigma_0=\sigma-M^3 A_3,\quad
    h_0=h-M A_1
\ee
where $\sigma$ and $h$ must be taken as an `input' parameters
to the theory in the spirit of \cite{Bordag 04}.
They describe the properties of material of the defect.
Just as electron mass $m$ and charge $e$ in QED,
values of $\sigma$ and $h$ cannot be predicted from the theory and must be
determined from appropriate experiments.
Thus, in addition to standard QED normalization conditions,
one needs three additional independent experiments to remove all the ambiguities of our
model ---
to determine $\sigma$ and $h$, and to set the scale of the coupling constant $a$.

\section{Conclusions}
We constructed the QED model which describes a photon field interacting with
semitransparent cylindrical shell (two-dimensional defect surface).
The form of interaction --- Chern-Simon action --- is completely determined by
the basic principles of QFT: locality, gauge and Lorenz invariance, renormalizability.
%
%
The defect action is parity odd and
P-transformation is equivalent to the change of sign of the defect coupling constant $a$.
We calculated explicitly the modified photon propagator and the Casimir energy.
The later one appears to be P-invariant being even function of $a$ and
tends with $a\to\infty$ to its value for perfectly conducting shell.
Parity violation manifests itself only when external field
is applied \cite{Markov Pismak'05}.
Thus, we can say that such defect action mimics the behavior of thin films
of magnitoelectrics.

Consideration of two-dimensional defects within the scope of renormalizable QFT is justified
by existing experimental data. It unambiguously shows that Casimir force for objects with
sharp boundaries has $1/r^3$ ($r$ --- distance between the objects)
behavior and thus is governed by dimensionless parameters of the system.
On the other hand, from obvious dimensional reasoning it follows that
the presence of any dimensional constants like finite width of the film $h$,
finite conductivity $\delta$, or final UV cut-off in non-renormalizable models,
can only give corrections in the order $h^2/r^5$, $\delta^2/r^5$ or higher. Thus, to the
next to leading order in inverse powers of $r$ we can stay with surface contributions only.

Effectively, with the defect action we model interaction of
electromagnetic field with a surface layer (thin film) of atoms constituting the media,
which can naturally be parity violating.
This property of the media translates into the surface properties
described in our model by a single constant $a$.
The parity in our model is preserved in two cases: the trivial one $a=0$, and
$a\to\infty$ that corresponds to perfect conductor limit as we showed above.
We predict that thin films which are parity even should have universal amplitude
of the Casimir force: either vanishing (with $a=0$),
or coinciding with one of the perfect conductor ($a=\infty$).
From theoretical point of view we can not decide whether
materials with finite defect coupling $a$ exist or not.


Presented calculations of the Casimir energy (and subsequently the Casimir force)
indirectly presume that the energy change is measured between two adiabatic
states of the system which differ by the radius of the cylinder. In particular this means that
if an experiment is
to be carried out, the cylinder must be deformed as a whole,
uniformly along all of its (infinite) length. Such experimental setup looks
unrealistic but still possible in principle. This rises arguments in the literature
(see review \cite{Milton 04})
that calcu\-lations of a renormalized (rigid) self-pressure has insignificant (if any) predictive power.

As a way to resolve contradictions,
one have to calculate `soft' self-pressure for local deformations of
the shape of the body and reveal its dependence on position along the body's surface.
%
%
%
For example,
consider a perfectly conducting cylinder shell of finite length $L$ and of radius $R$.
%
%
%
%
It is clear that in the limit $R/L\to\infty$ in a vicinity of the cylinder axis
the local self-pressure must reproduce the attractive Casimir force
between parallel plates --- the two bases of the cylinder.
On the other hand, when $R/L\sim 1$
one can consider the deformations of the cylinder as a whole, and should recover the rigid stress.
We call such system a Casimir drum. Explicit calculation of the Casimir energy, rigid and soft
self-pressures of Casimir drum is the next step for our research.
%
%
%
%
\section*{Acknowledgement}

Authors would like to thank Prof R.~Jaffe and Prof D.~Vassilevich for fruitful discussions on the
subject of the paper.
V.N. Markov and Yu.M. Pismak are grateful to Russian Foundation of Basic Research
for financial support (RFRB grant $07$--$01$--$00692$).

\end{document}